\pdfoutput=1
\documentclass[11pt,twoside,a4paper,cmspaper,final,collab]{cms-tdr}
\def\svnVersion{fe592cf}\def\svnDate{2023/07/28}\def\cmsCernNoTag{CERN-EP-2022-233}\def\cmsCernDate{\today}\def\cmsMessage{Published in Physical Review Letters as \href{http://dx.doi.org/10.1103/PhysRevLett.131.041801}{\doi{10.1103/PhysRevLett.131.041801}.}}
\begin{document}\cmsNoteHeader{HIG-21-012}

\newlength\cmsTabSkip
\setlength{\cmsTabSkip}{1ex}
\newlength\cmsFigWidth
\if@twocolumn
  \setlength{\cmsFigWidth}{0.96\columnwidth}
\else
  \setlength{\cmsFigWidth}{0.48\textwidth}
\fi
\ifthenelse{\boolean{cms@external}}{\providecommand{\cmsLeft}{upper\xspace}}{\providecommand{\cmsLeft}{left\xspace}}
\ifthenelse{\boolean{cms@external}}{\providecommand{\cmsRight}{lower\xspace}}{\providecommand{\cmsRight}{right\xspace}}
\providecommand{\cmsTable}[1]{\resizebox{\textwidth}{!}{#1}}

\newcommand{\GeVnbs}{\ensuremath{\text{Ge\hspace{-.08em}V}}}
\newcommand{\hbb}{\ensuremath{\PH\to\bbbar}\xspace}
\newcommand{\hcc}{\ensuremath{\PH\to\ccbar}\xspace}
\newcommand{\hzbb}{\ensuremath{\PH(\PZ)\to\bbbar}\xspace}
\newcommand{\hzcc}{\ensuremath{\PH(\PZ)\to\ccbar}\xspace}
\newcommand{\zcc}{\ensuremath{\PZ\to\ccbar}\xspace}
\newcommand{\zbb}{\ensuremath{\PZ\to\bbbar}\xspace}
\newcommand{\xcc}{\ensuremath{\PX\to\ccbar}\xspace}
\newcommand{\Vjets}{\ensuremath{\PV}+jets\xspace}
\newcommand{\mH}{\ensuremath{m_{\PH}}\xspace}
\newcommand{\mSD}{\ensuremath{m_{\mathrm{SD}}}\xspace}
\newcommand{\nddt}{\ensuremath{N_{2}^{1\mathrm{,DDT}}}\xspace}
\newcommand{\R}{\ensuremath{R}\xspace}
\newcommand{\Rpf}{\ensuremath{R_{\mathrm{p}/\mathrm{f}}}\xspace}
\newcommand{\HJMINLO}{{\textsc{HJ-MiNLO}}\xspace}
\newcommand{\MINLO}{{\textsc{MiNLO}}\xspace}
\newcommand{\fxfx}{{\textsc{FxFx}}\xspace}
\newcommand{\ggF}{\ensuremath{\Pg\Pg\mathrm{F}}\xspace}
\newcommand{\muObsLimit}{\ensuremath{47}\xspace}
\newcommand{\muExpLimit}{\ensuremath{39}\xspace}
\newcommand{\muHObsVal}{\ensuremath{9.4}\xspace}
\newcommand{\muHObsErr}{\ensuremath{_{-19.9}^{+20.3}}\xspace}
\newcommand{\muZVal}{\ensuremath{1.00}\xspace}
\newcommand{\muZErrUp}{\ensuremath{0.19}\xspace}
\newcommand{\muZErrDown}{\ensuremath{0.17}\xspace}
\newcommand{\muZStatErr}{\ensuremath{0.06}\xspace}
\newcommand{\muZThyErr}{\ensuremath{0.08}\xspace}
\newcommand{\muZSystErr}{\ensuremath{_{-0.14}^{+0.17}}\xspace}
\newcommand{\muZline}{\ensuremath{\muZVal\muZSystErr\syst\pm\muZThyErr\thy\pm\muZStatErr\stat}\xspace}
\newcommand{\muZlineshort}{\ensuremath{\muZVal_{-\muZErrDown}^{+\muZErrUp}}\xspace}
\newcommand{\sigmaH}[1][{}]{\ensuremath{\sigma_{\text{#1}}\left(\PH\right)}\xspace}
\newcommand{\brHCC}[1][{}]{\ensuremath{\mathcal{B}_{\text{#1}}\left(\PH\to\PQc\PAQc\right)}\xspace}
\newcommand{\sigmaB}[1][{}]{\ensuremath{\sigmaH[#1]\brHCC[#1]}\xspace}

\newcommand{\analysisLumi}{\ensuremath{138\fbinv}\xspace}

\newcommand{\DeepCSV}{{\textsc{DeepCSV}}\xspace}
\newcommand{\DeepDoubleX}{{\textsc{DeepDoubleX}}\xspace}
\newcommand{\DeepDoubleBvL}{{\textsc{DeepDoubleBvL}}\xspace}
\newcommand{\DeepDoubleCvL}{{\textsc{DeepDoubleCvL}}\xspace}
\newcommand{\DeepDoubleCvB}{{\textsc{DeepDoubleCvB}}\xspace}
\newcommand{\DDX}{{\textsc{DDX}}\xspace}
\newcommand{\DDBvL}{{\textsc{DDBvL}}\xspace}
\newcommand{\DDCvL}{{\textsc{DDCvL}}\xspace}
\newcommand{\DDCvB}{{\textsc{DDCvB}}\xspace}

\renewcommand{\PYTHIA} {{\textsc{Pythia}}\xspace}
\renewcommand{\POWHEG} {{\textsc{POWHEG}}\xspace}

\cmsNoteHeader{HIG-21-012}
\title{Search for Higgs boson and observation of Z boson through their decay into a charm quark-antiquark pair in boosted topologies in proton-proton collisions at \texorpdfstring{$\sqrt{s} = 13\TeV$}{sqrt(s)=13 TeV}}

\date{\today}

\abstract{A search for the standard model (SM) Higgs boson (\PH) produced with transverse momentum (\pt) greater than 450\GeV and decaying to a charm quark-antiquark (\ccbar) pair is presented. The search is performed using proton-proton collision data collected at $\sqrt{s}=13\TeV$ by the CMS experiment at the LHC, corresponding to an integrated luminosity of 138\fbinv. Boosted \hcc decay products are reconstructed as a single large-radius jet and identified using a deep neural network charm tagging technique. The method is validated by measuring the \zcc decay process, which is observed in association with jets at high \pt for the first time with a signal strength of $1.00_{-0.14}^{+0.17}\syst\pm0.08\thy\pm0.06\stat$, defined as the ratio of the observed process rate to the SM expectation. The observed (expected) upper limit on \sigmaB is set at 47 (39) times the SM prediction at 95\% confidence level.}

\hypersetup{
  pdfauthor={CMS Collaboration},
  pdftitle={Search for boosted Higgs boson decay to a charm quark-antiquark pair in proton-proton collisions at sqrt(s)=13 TeV},
  pdfsubject={CMS},
  pdfkeywords={CMS, Higgs, charm, gluon fusion, Z, DeepDoubleX}
}

\maketitle
The standard model (SM) Higgs boson (\PH) has been observed at the
LHC~\cite{:2012gk,:2012gu,Chatrchyan:2013lba} in all its expected primary
production modes and most of its dominant decay channels. With the observations
of direct couplings to \PGt leptons~\cite{Sirunyan:2017khh,ATLAS:2022yrq}, top
quarks~\cite{Sirunyan:2018hoz,ATLAS:2018mme}, and bottom
quarks~\cite{ATLAS:2018kot,Sirunyan:2018kst} confirming that the SM Yukawa
sector gives rise to the masses of third-generation fermions, attention
naturally turns to probing the second generation, specifically muons and charm
quarks.

The search for \PH decays to muon pairs is the most experimentally accessible
channel, and has been explored by the ATLAS~\cite{Sirunyan:2018hbu,2021135980}
and CMS~\cite{Aaboud:2017ojs} Collaborations. The latter recently found evidence of
the \PH coupling to muons~\cite{cmscollaboration2020evidence}.

In contrast, the search for \PH decays to charm quark-antiquark pairs (\hcc) is
considerably more challenging, because of the difficulty of identifying such
decays and due to the enormous multijet background. However, recent advances in
jet substructure and flavor tagging techniques~\cite{CMS-DP-2018-046} have
greatly improved the experimental sensitivity to this decay mode. Prior searches
by ATLAS~\cite{Aaboud_2018,ATLAS:2022ers} and
CMS~\cite{Sirunyan:2019qia,CMS:2022psv} have focused on \PH production in
association with a vector boson ($\PV\PH$, where \PV stands for a \PW or \PZ
boson), which benefits from strong background rejection thanks to leptonic
decays of the vector bosons. The gluon-gluon fusion (\ggF) and vector boson
fusion (VBF) production modes have larger cross sections, but have yet to be
explored.

This letter reports on the first search for the \hcc decay at the LHC,
where the \PH  is produced with transverse momentum (\pt) greater than $450\GeV$,
enriched in events from the \ggF production. The search employs the same general
strategy as earlier CMS searches for boosted \PH in the \bbbar decay
channel~\cite{Sirunyan:2017dgc,CMS:2020zge}, but uses new mass decorrelated
discriminators to define a charm-enriched signal region. This search strategy
provides an additional constraint on the decay process to the existing ATLAS and
CMS measurements, in terms of production mode and \PH \pt.

The search is performed using a dataset of proton-proton collisions at $\sqrt{s}
= 13\TeV$, collected with the CMS detector at the LHC, and corresponding to an
integrated luminosity of 138\fbinv. Candidate events are selected by requiring a
high-\pt, large-radius jet with substructure observables compatible with those
expected from an \hcc decay. Deep neural network (DNN) discriminators are
employed to separate the \PH signal events from the dominant background,
specifically quantum chromodynamics (QCD)-induced multijet events. The
discriminators are designed to be independent of jet mass, which allows for both
an estimation of the QCD background from control samples in data and for the
validation of the analysis procedure through a search for \zcc decays. A
model of the jet mass distributions for the \hcc and \zcc signals, QCD multijet events, and
other background processes is fit simultaneously in several disjoint signal and
control regions to extract the signal production cross sections.

The CMS apparatus~\cite{Chatrchyan:2008zzk} is a multipurpose, nearly hermetic detector,
designed to trigger on~\cite{Khachatryan:2016bia,CMS:2020cmk} and identify electrons,
muons, photons, and (charged and neutral)
hadrons~\cite{CMS:2020uim,Sirunyan:2018fpa,Chatrchyan:2014fea}. A global
``particle-flow" (PF) algorithm~\cite{Sirunyan:2017ulk} aims to reconstruct all
individual particles in an event, combining information provided by the all-silicon
inner tracker and by the crystal electromagnetic and brass-scintillator hadron
calorimeters, operating inside a 3.8\unit{T} superconducting solenoid, with data from
the gas-ionization muon detectors embedded in the flux-return yoke outside the solenoid.
The reconstructed particles are used to build \PGt leptons, jets, and missing transverse
momentum ($\ptmiss$)~\cite{CMS:2018jrd,Khachatryan:2016kdb,CMS:2019ctu}.

{\tolerance=800 
Simulated samples of signal and background events are produced at the matrix
element level using various Monte Carlo (MC) event generators. The QCD multijet,
$\PZ$+jets, and $\PW$+jets processes are modeled at QCD leading order (LO) accuracy
using the \MGvATNLO~v2.4.2 generator~\cite{Alwall:2014hca}. The vector boson
samples contain decays to all flavors of quarks, and include up to 3 (4) extra
partons at the matrix element level for \Vjets events. Jets from the
matrix element calculations and the parton shower description are matched using
the MLM prescription~\cite{Alwall:2007fs}. The \ttbar and single top quark
processes are modeled at next-to-LO (NLO) using
\POWHEG~2.0~\cite{Nason:2004rx,Frixione:2007nw,Frixione:2007vw,Alioli:2010xd,Re:2010bp,Frederix:2012dh}.
The diboson processes are modeled at LO accuracy with
\PYTHIA8.226~\cite{Sjostrand:2014zea}. For 2016 data-taking conditions, QCD
samples are modeled with \MGvATNLO~v2.2.2, and diboson samples are modeled with
\PYTHIA8.205.
\par}

The differential cross sections for the $\PZ$+jets and $\PW$+jets (\Vjets)
samples are corrected with boson \pt-dependent functions for higher-order QCD
and electroweak (EW) effects. The QCD NLO corrections are derived using
\MGvATNLO, simulating \PW and \PZ production with up to two additional partons
and the \fxfx matching to the parton shower~\cite{Frederix:2012ps}. The EW NLO
corrections are taken from theoretical calculations of
Refs.~\cite{Kallweit:2014xda,Kallweit:2015fta,Kallweit:2015dum,Lindert:2017olm}.
Additionally, the total cross sections for the diboson samples are corrected to
next-to-NLO (NNLO) accuracy with the \MCFM~7.0 program~\cite{Campbell:2010ff}.

The \ggF \PH production process is simulated using the {\HJMINLO}~\cite{Frixione:2007vw,Hamilton:2012rf,Luisoni:2013kna,Becker:2669113}
event generator with mass $\mH = 125\GeV$ and including finite top quark mass
effects, following the recommendation in Ref.~\cite{Becker:2669113}. The
\POWHEG~\cite{Bagnaschi2012} generator is used to model the \PH production
through the VBF, $\PV\PH$, and $\ttbar\PH$
processes~\cite{Nason:2009ai,Luisoni:2013kna,Hartanto:2015uka}.

For parton showering and hadronization, the \POWHEG and \MGvATNLO samples are interfaced
with \PYTHIA8.205 (8.230) for 2016 (2017--2018) running conditions. The corresponding
\PYTHIA parameters for the underlying event description are set to the
CUETP8M1~\cite{Khachatryan:2015pea} (CP5~\cite{Sirunyan:2019dfx}) tune. For 2016 samples
the parton distribution function (PDF) set {NNPDF3.0}~\cite{Ball:2014uwa} is used, with
either LO or NLO accuracy, corresponding to that used in the matrix element
calculations, while for 2017--2018 samples {NNPDF3.1}~\cite{Ball:2017nwa} at NNLO
accuracy is used for all processes. The detector response is modeled with
\GEANTfour~\cite{GEANT4}.

Reconstructed particles are clustered into jets using the anti-\kt
algorithm~\cite{Cacciari:2008gp,Cacciari:2011ma}. Small-radius jets are clustered with a
distance parameter of 0.4 (AK4 jets). Large-radius jets arising from the decays of
boosted heavy particles are reconstructed with a distance parameter of 0.8 (AK8
jets). The effect of particles from additional proton-proton interactions within the same or
nearby bunch crossings (pileup) is mitigated through the charged hadron
subtraction~\cite{Sirunyan:2017ulk} and pileup-per-particle
identification~\cite{Bertolini:2014bba} algorithms for AK4 and AK8 jets, respectively.
Additional corrections are applied to the jet energy as functions of jet pseudorapidity
($\eta$) and \pt to account for the detector response.

The \PH candidate is reconstructed as a single AK8 jet with $\pt > 450\GeV$. A mix of
triggers using jet \pt or a scalar sum of the jet \pt in the event is employed for
online selection. At $\pt = 450 \GeV$, the online selection is 90\% efficient with
respect to the offline selection, reaching full efficiency by 500\GeV. The soft-drop
(SD) algorithm~\cite{Larkoski:2014wba} with parameters $\beta=0$ and $z=0.1$ is applied
to the jet mass (\mSD) to remove soft and wide-angle radiation, which reduces the mass
of jets originating from QCD background events while preserving the mass of jets
originating from heavy boson decays. The range of interest is set to $40 < \mSD <
  201\GeV$. To match the tracker acceptance region, jets are required to have $\abs{\eta}
  < 2.5$. In case of several jets in the event passing the criteria, the jet with the
highest charm versus light tagging score, defined below, is taken to be the \PH candidate.

After the above selections, the dominant background is the QCD multijet
production, which accounts for more than $95\%$ of the expected yield and is
estimated from data. The \Vjets processes are significant resonant backgrounds
at approximately $4\%$, and are estimated from simulation. The \ttbar process
constitutes a subdominant nonresonant background across the \mSD spectrum, the
shape of which is taken from simulation, while the total yield is estimated from
data. Other EW processes, including diboson, triboson, and $\ttbar\PV$
processes, are estimated from simulation and found to be negligible.

Events containing leptons are vetoed to reduce SM EW backgrounds. The selection criteria
for electrons, muons, and hadronic \PGt leptons are $\pt > 10$, 10, and 20\GeV and
$\abs{\eta} < 2.5$, 2.4, and 2.3, along with ``veto'', ``loose'', and ``very loose''
identification requirements~\cite{CMS:2020uim,CMS:2019ied,CMS:2018jrd}, respectively. In
addition, muons are required to have a relative isolation (scalar \pt sum of the PF
candidates within a cone with a distance parameter of 0.4, divided by the lepton \pt) of
less than 0.25. Events with $\ptmiss > 140 \GeV$, as well as events with AK4
\cPqb-tagged jets with $\pt > 30\GeV$ opposite in azimuth to the \PH candidate jet
($\Delta\phi(\text{AK4},\text{AK8}) > \pi/2$), are removed to reduce the top quark
background. The AK4 \cPqb jet identification is performed using the \DeepCSV DNN
algorithm~\cite{Sirunyan:2017ezt} with a working point corresponding to a 1\%
misidentification probability for light (\PQu, \PQd, \PQs quark, or gluon) jets.

The dimensionless mass scale variable
$\rho=2\ln(\mSD/\pt)$~\cite{Dasgupta:2013ihk,Dolen:2016kst} is used to parametrize the
QCD background model (described below) as its distribution is approximately invariant
versus jet \pt, unlike jet \mSD. A selection of $-6.0 < \rho < -2.1$ is imposed to avoid
instabilities and edge effects from the SD algorithm and jet
clustering~\cite{ATLAS:2019mgf}. The lower $\rho$ threshold implies an upper jet $\pt$
threshold, which is made explicit by requiring $\pt < 1200\GeV$. In simulation, less
than 1\% of signal events are found above this upper bound.

The $N_2^{1}$ variable~\cite{Moult:2016cvt}, a ratio of energy correlation
functions~\cite{Larkoski:2013eya}, is a powerful way to identify two-pronged signatures.
However, using it for selection distorts the background jet mass distribution as a
function of \pt. To mitigate this effect, the designing decorrelated taggers (DDT)
technique~\cite{Dolen:2016kst}, effectively a sliding selection, is applied. The
selection is on $\nddt \equiv N_2^{1} - q_{0.26}(\pt,\rho) < 0$, where
$q_{0.26}(\pt,\rho)$ is the $N_2^{1}$ value corresponding to the 26\% efficiency for the
QCD background, as a function of jet \pt and $\rho$. The target percentile is chosen to
optimize the \hcc expected significance.

\begin{figure}[hbtp]
  \centering
  \includegraphics[width=\cmsFigWidth]{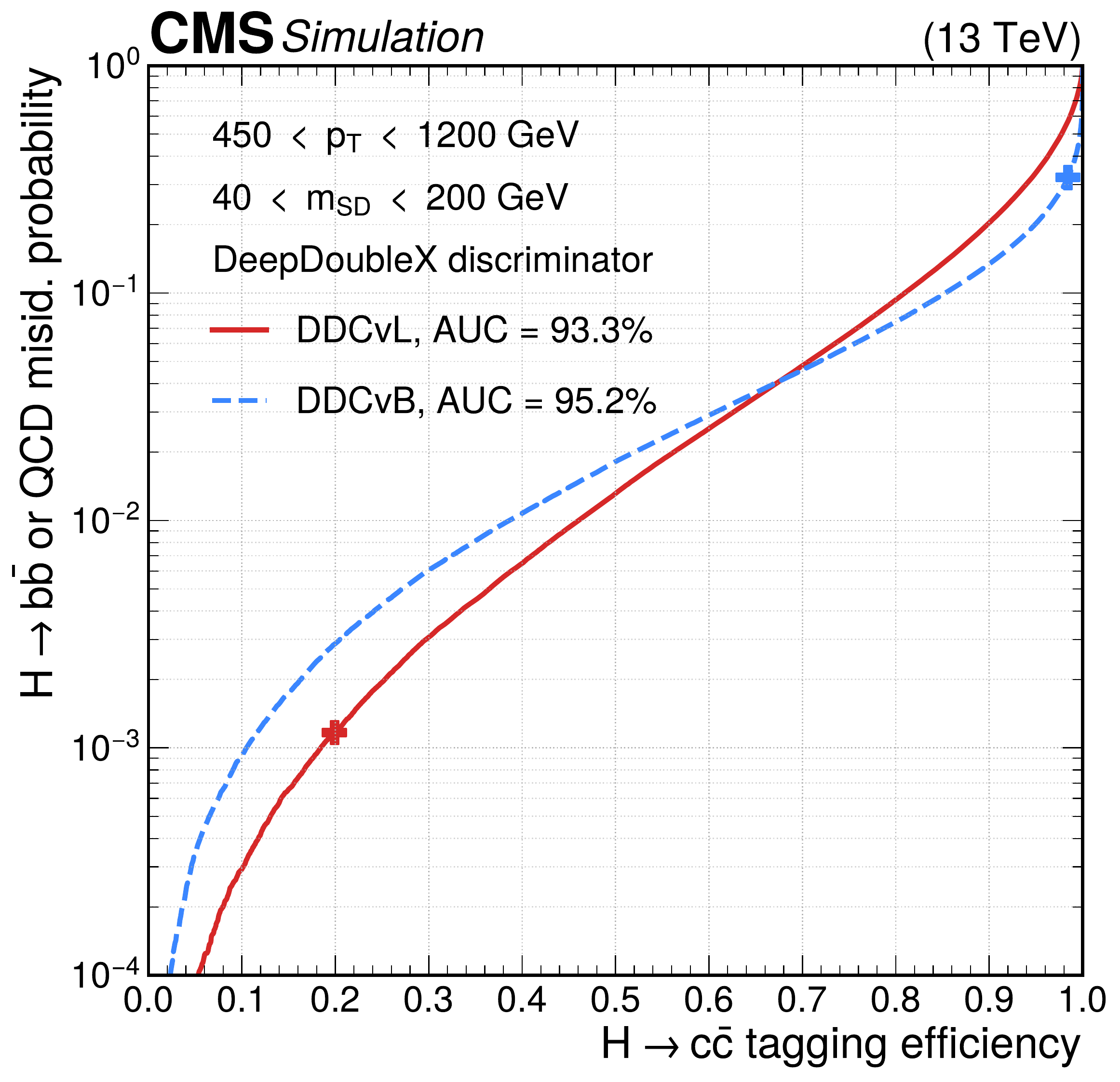}
  \caption[\DeepDoubleX]{The \DDCvL and \DDCvB performance for \hcc identification versus
    QCD and \hbb processes, respectively. No selection apart from the displayed
    \mSD and \pt requirements is applied. The working points used in this search
    are marked with a cross. The AUC is the area-under-curve metric.
  }
  \label{fig:ROCDDX}
\end{figure}

Finally, jet flavor is determined by the \DeepDoubleX DNN
algorithm~\cite{CMS-DP-2022-041}. The model comprises convolutional and
recurrent units processing low-level features of secondary vertices and PF
candidates, the outputs of which are joined with expert
variables~\cite{Sirunyan:2017ezt} in a fully connected layer. The application of
feature importance ranking techniques, such as integrated
gradients~\cite{integrated} and deep Taylor
decomposition~\cite{MONTAVON2017211}, indicates the key features to be the
angular distances of the PF candidates from both the jet and
2-subjetiness~\cite{Thaler:2010tr} axes. The kinematic properties of the PF
candidates defined relative to the parent jet have subleading importance. The
model is trained to distinguish between two-pronged \PH-like signatures of
bottom and charm flavors, as well as the QCD background, yielding two per-jet
classifiers: charm versus light, referred to as \DeepDoubleCvL (\DDCvL),
and charm versus bottom, referred to as \DeepDoubleCvB (\DDCvB). The
performance of the two classifiers is shown, prior to any analysis-specific
selection, in Fig.~\ref{fig:ROCDDX}. The optimal working point, maximizing the
\hcc expected significance after all previous selections are applied, is found
with respect to both classifiers at a QCD efficiency of 0.5\% and a \hcc
efficiency of 20.6\%. The corresponding efficiency for \hbb events is 4.8\%. As
the classifiers are mass independent, the quoted efficiencies also apply to \zcc and
\zbb.

The relative contributions of \PH production modes to the overall signal yield
are 55\%, 25\%, and 20\% for \ggF, VBF, and $\PV\PH$, respectively, and are
similar for both \ccbar and \bbbar decay modes. The $\ttbar\PH$ contribution is
suppressed due to the top veto and found to be negligible. Events passing all of
the selection requirements described above constitute the signal or ``passing''
region (SR), whereas events failing the DDCvL requirement while passing the
rest, including the DDCvB requirement, constitute the control or ``failing''
region (CR). Both the SR and CR are subdivided into 23 evenly spaced bins of jet
\mSD in the range 40--201\GeV and 6 \pt bins from 450 (or 475) to 1200\GeV.
Additionally, all regions are subdivided according to the three data-taking
years (2016--2018).

The QCD background is not accurately predicted in simulation. Since the flavor
discrimination is nearly independent of the jet \pt and mass, the ratio of the
passing and failing region distributions, \Rpf, is expected to be approximately
flat with respect to the jet \pt and \mSD. This can be exploited to obtain an SR
prediction of the QCD background from the CR via an appropriate efficiency
scaling. A residual difference in shapes can be accounted for by parametrizing
the \Rpf shape in the two dimensions. In order to take into account a potential
mass dependence of the flavor selection efficiency, a correction factor,
$\Rpf^{\mathrm{QCD}}(\pt,\rho)$, is fit to the simulated QCD background shapes.
Then, a second correction factor of the same functional form,
$R^{\text{data}}(\pt,\rho)$, accounts for mismodellings in simulation. Both
are parametrized in terms of Bernstein polynomials~\cite{bernstein1912} in \pt
and $\rho$:
\begin{equation}
  \R(\pt, \rho) = \sum^{n_{\rho}}_{k=0} \sum^{n_{\pt}}_{\ell=0} a_{k, \ell} b_{\ell, n_{\pt}}(\pt) b_{k, n_{\rho}}(\rho),
\end{equation}
where $n_{\pt}$ is the degree of the polynomial in $\pt$, $n_{\rho}$ is the degree of
the polynomial in $\rho$, $a_{k, \ell}$ is a Bernstein coefficient, and $b_{\nu, n}$ is
a Bernstein basis polynomial of degree $n$. The coefficients $a_{k, \ell}$ of
$\Rpf^{\mathrm{QCD}}(\pt,\rho)$ are determined in a fit to simulated QCD background
events; $a_{k, \ell}$ of $\Rpf^{\text{data}}(\pt,\rho)$ are unconstrained and are
determined during the maximum likelihood fit to data. The total effective \Rpf is then
expressed as:
\begin{equation}
  \Rpf(\pt,\rho) = \Rpf^{\mathrm{QCD}}(\pt,\rho) R^{\text{data}}(\pt,\rho).
\end{equation}

The $\Rpf$ are expected to vary between data-taking years because of changes in detector
conditions, and are thus fit independently for each year. The minimal degree of the
polynomials necessary to fit the QCD simulation and the data is determined by a Fisher
F-test~\cite{ref:ftest} and found to be $(n_{\pt},\,n_{\rho})=(0,\,2), (1,\,2)$, and
$(0,\,2)$ for $\Rpf^{\mathrm{QCD}}(\pt,\rho)$ and $(1,\,0), (0,\,0)$, and $(1,\,0)$ for
$R^{\text{data}}(\pt,\rho)$ for the years 2016, 2017, and 2018, respectively. Bias
tests were performed with respect to the choice of parametrization, and no significant
bias was found.

The \Vjets processes are modeled using simulation. The differential \ttbar
contribution is taken from simulation; however, the normalizations in the SR and
CR are corrected via two scale factors measured in a dedicated \ttbar-enriched
control region, parametrizing the overall normalization and the efficiency of
the DDCvL selection between the SR and CR. This control region is adapted from
the SR selection by lowering the \PH candidate \pt threshold, requiring exactly
one muon, and inverting the selection requirements on \ptmiss and $\PQb$-tagged
AK4 jets. The scale factor measurement is performed \emph{in situ} during the
signal extraction, separately for each data-taking period, and the values are
given in Table~\ref{tab:SF}. The \hbb contribution is taken from the simulation
and is fixed to the SM expectation. While its expected SR yield is greater by
approximately a factor of 5 than that of the \hcc signal, its impact is
negligible with respect to the overall background uncertainty.

The dominant systematic uncertainties for this search are related to the flavor tagging
efficiency and jet mass shape. Corrections of the jet mass, jet mass resolution, and
\nddt and DDCvB efficiencies are derived from data using \PW boson jets from
semileptonic \ttbar events. These corrections are measured independently of jet flavor,
and as such are correlated among all considered resonant (\PH, \PZ, \PW) production and
decay processes. The DDCvL misidentification efficiency of the \PW process is measured
here as well.
The corrections and their associated uncertainties are given in Table~\ref{tab:SF}.

\begin{table*}[htb!]
  \centering
  \topcaption{Summary of the applied data-to-simulation scale factors for the jet mass,
    jet mass resolution, $\nddt$ selection, \ttbar normalization, and \DeepDoubleX selections for
    different data-taking periods. The jet mass correction is additive, in units of
    \GeV.}
    \cmsTable{
    \begin{scotch}{cccccccc}
      Data   & Jet mass             & Jet mass           & $\nddt$,         & \ttbar             & CvL selection   & CvL selection   & CvL selection \\
      period & correction [\GeVnbs] & resolution         & CvB selection    & normalization      & \ttbar          & ($\PW$+jets)    & (signal)        \\
      \hline
      2016   & $ -1.17 \pm 0.22$    & $ 1.021 \pm 0.017$ & $ 0.89 \pm 0.02$ & $ 0.84 \pm 0.05$ & $ 0.93 \pm 0.15$  & $0.62 \pm 0.09$ & $1.15 \pm 0.25$ \\
      2017   & $ -1.19 \pm 0.23$    & $ 1.019 \pm 0.016$ & $ 0.90 \pm 0.02$ & $ 0.86 \pm 0.09$ & $ 0.93 \pm 0.15$  & $0.64 \pm 0.09$ & $0.85 \pm 0.16$ \\
      2018   & $ -0.12 \pm 0.21$    & $ 1.090 \pm 0.031$ & $ 0.92 \pm 0.02$ & $ 0.86 \pm 0.08$ & $ 1.00 \pm 0.14$  & $0.72 \pm 0.08$ & $0.74 \pm 0.20$ \\
    \end{scotch}
    }
  \label{tab:SF}
\end{table*}

The efficiency of the \textsc{DDCvL} selection for the signal processes is estimated
using data and simulation samples enriched in $\ccbar$ pairs from gluon
splitting~\cite{Sirunyan:2017ezt}. Signal-like events are selected by requiring each of
the two SD subjets of an AK8 jet to contain a muon, targeting semileptonic decays of
\PQb/\PQc hadrons. The efficiency is extracted from a template fit to the combined mass
of all matched secondary vertices; the measured correction factors are given in
Table~\ref{tab:SF}.
The relative uncertainty of the misidentification efficiency of $\bbbar$ decays is assigned to be 30\%. 
Varying this value from 10 to 50\% has a negligible effect on reported results.

Other systematic uncertainties are assigned to cover potential mismodeling of the \PH
signal, in particular for the \ggF and VBF production modes~\cite{Becker:2669113}, and
higher-order corrections to the \PW and \PZ processes~\cite{Lindert:2017olm}. Finally,
systematic uncertainties for experimental effects, including jet energy scale and
resolution~\cite{CMS:jecs}, trigger and veto efficiencies~\cite{CMS:btv16,CMS:btv1718},
variations in the measured pileup~\cite{CMS:pileup}, finite simulated sample
size~\cite{Barlow:1993dm}, and an integrated luminosity
measurement~\cite{CMS:lumi,CMS:lumi17,CMS:lumi18}, are also included, but are found to
have a comparatively small effect.

The parameter of interest in this analysis is the signal strength $\mu_\PH$ or
$\mu_\PZ$, defined as the ratio of the observed to the SM expected \PH or \PZ
boson production cross section times the \hcc or \zcc branching fraction,
respectively. These parameters are extracted from a binned (\mSD, \pt) maximum
likelihood fit to the observed data, where the expected value is the sum of the
signal contribution (scaled by the signal strength parameter) and the background
contributions, each modified by nuisance parameters to account for the
previously discussed systematic effects. The magnitude of each systematic
uncertainty is encoded in the likelihood model as an additional constraint,
treated according to the frequentist paradigm~\cite{LHCCLs}. The fit is
performed simultaneously across all subdivisions of the SR and CR described
previously, as well as the per-year \ttbar background enriched CRs.

To validate the analysis strategy, and to confirm the presence of \zcc decays,
the \PZ signal strength $\mu_\PZ$ is measured via a profile likelihood fit, treating
$\mu_\PH$ as a nuisance parameter, and is found to be \muZline. This corresponds to an
excess, both observed and expected, over the $\mu_\PZ=0$ hypothesis with a significance
of well over $5$ standard deviations.
The precision of the $\mu_\PZ$ measurement is primarily limited by the systematic
uncertainty in the DDCvL signal tagging efficiency. The subleading uncertainty comes
from the modeling of the $\PZ$+jets production cross section.

For the extraction of $\mu_{\PH}$, since the $\PZ$ cross section has been measured in
leptonic decay channels and found to agree with theoretical predictions within 5\% in
this \pt regime~\cite{CMS:2020hkd} and since the \zcc branching
ratio is known to 2\% precision~\cite{Workman:2022ynf}, we fix $\mu_\PZ \equiv 1$,
constraining the expected \PZ contribution to be within the applicable uncertainties of
its SM value. This serves to further constrain \emph{in situ} the DDCvL signal tagging
efficiency uncertainty. The measured efficiencies are compatible with the values quoted
in Table~\ref{tab:SF} and have approximately 30\% lower uncertainty.

\begin{figure}[ht!]
  \centering
  \includegraphics[width=\cmsFigWidth]{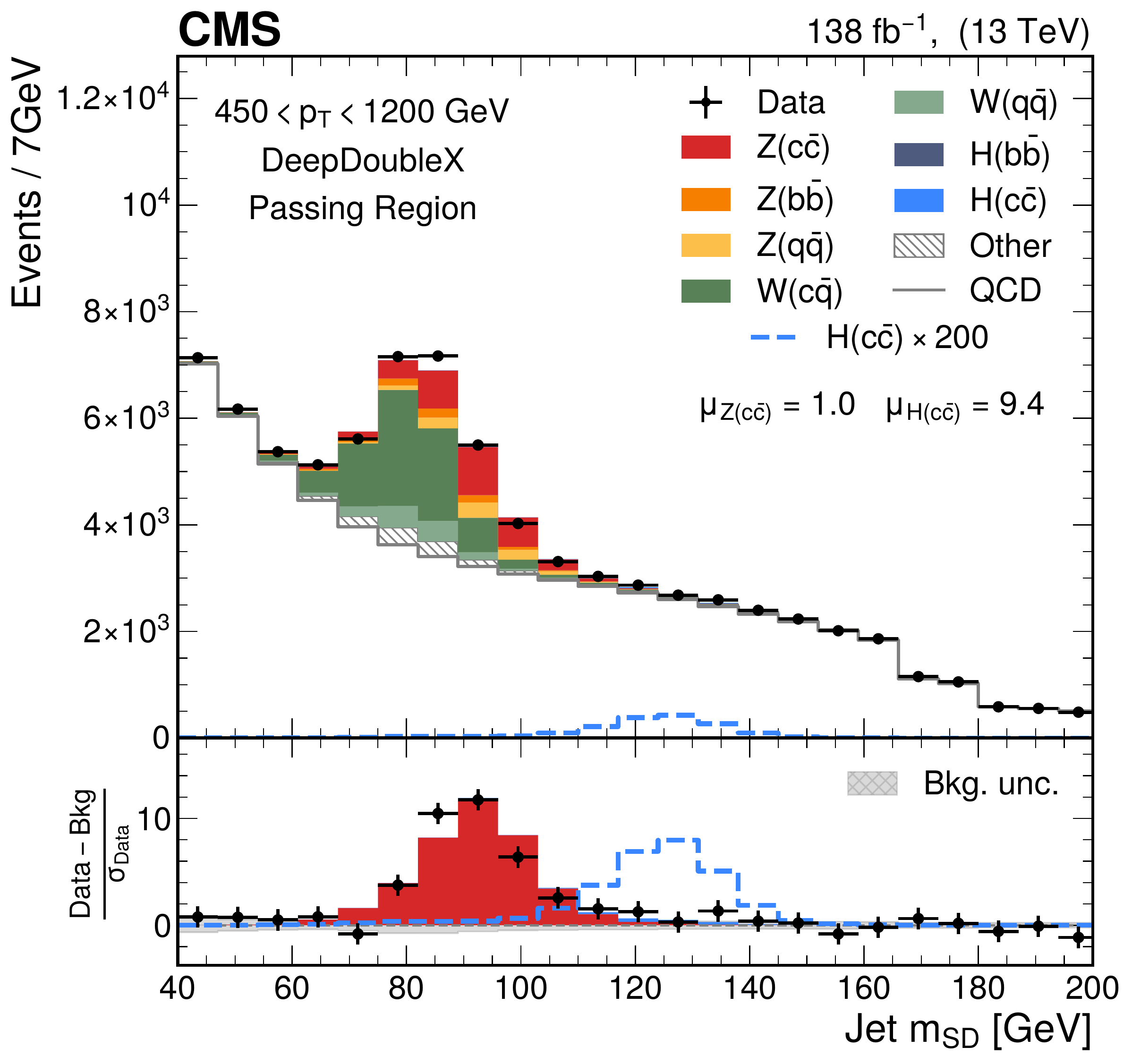}
  \includegraphics[width=\cmsFigWidth]{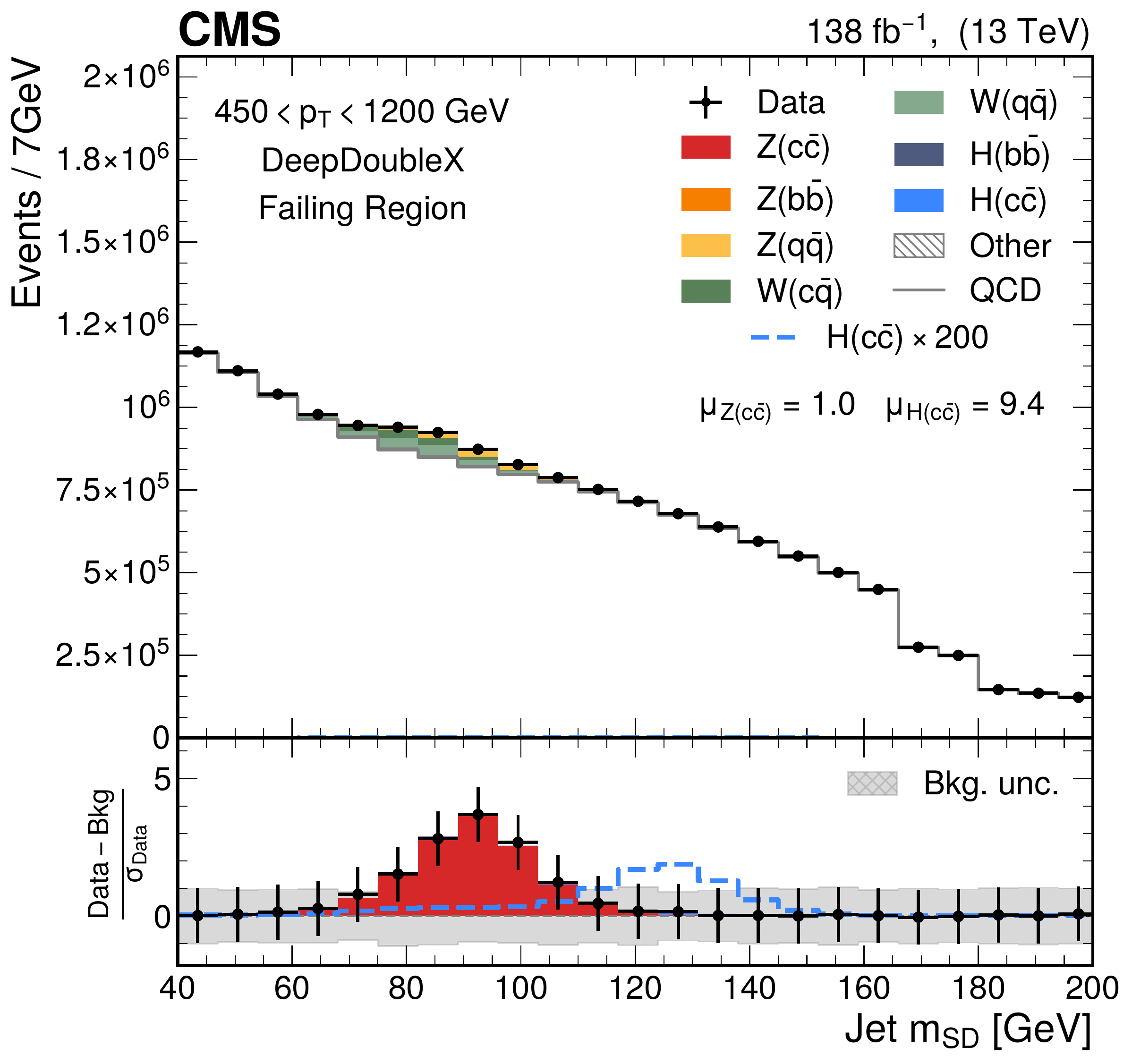}
  \caption{The observed and fitted \mSD distributions for the passing (\cmsLeft)
    and failing (\cmsRight) regions, combining all \pt categories and the three
    data-taking years. The fit is performed under the signal-plus-background
    hypothesis with a~single inclusive \hcc signal strength parameter.
    The \ttbar background yields and the QCD background yields and shapes are
    estimated from data. The \ttbar process constitutes the majority of
    contributions labeled ``other''. The dashed line represents the \hcc
    expectation, multiplied by a factor of 200. The steplike features at 166
    and 180\GeV are due to the $\rho$ acceptance upper bound, which excludes
    bins with high \mSD and low \pt from the analysis. The lower panel shows the
    residual difference between the data and the overall background (excluding
    \zcc), divided by the statistical uncertainty in the data. The near perfect
    model agreement with data in the failing region (\cmsRight) is by
    construction.}
  \label{fig:mSD}
\end{figure}

An observed (expected) upper limit is placed on the signal strength $\mu_\PH$ using the
profile likelihood ratio test statistic~\cite{LHCCLs}, \CLs criterion~\cite{CLS2,CLS1},
and asymptotic formulae~\cite{Cowan:2010js}, and found to be $\muObsLimit$
($\muExpLimit$) at 95\% confidence level. For the best fit value of
$\mu_\PH=\muHObsVal\muHObsErr$, the total \mSD distributions in the passing and
failing regions are shown in Fig.~\ref{fig:mSD}, and a breakdown of the sources of
uncertainty affecting the measurement is shown in Table~\ref{tab:systematics}. Tabulated
results are provided in the HEPData record for this analysis~\cite{hepdata}.

\begin{table}[ht!]
  \newcolumntype{.}{D{.}{.}{3.1}}
  \centering
  \topcaption{Sources of uncertainty in the measurement of the signal strength
    $\mu_{\PH}=\muHObsVal\muHObsErr$, and their observed impact ($\Delta\mu_{\PH}$) in
    the fit to the full data set. The impact of each uncertainty is evaluated by
    computing the uncertainty excluding that source and subtracting it in quadrature
    from the total uncertainty. The total uncertainty does not match the sum in
    quadrature of each source because of correlations among the components.}
  \begin{scotch}{ l .. }
    Uncertainty source                                & \multicolumn{2}{c}{$\Delta\mu_{\PH}$}           \\
    \hline
    Statistical                                       & $+16.7$                               & $-16.6$ \\
    \qquad Signal extraction                          & $+14.2$                               & $-14.1$ \\
    \qquad QCD pass-fail ratio (data correction)      & $+7.4$                                & $-7.4$  \\
    \qquad \ttbar normalization and misidentification & $+0.9$                                & $-0.7$  \\
    [\cmsTabSkip]
    Systematic                                        & $+10.5$                               & $-10.4$ \\
    \qquad QCD pass-fail ratio (simulation)           & $+9.7$                                & $-9.8$  \\
    \qquad Flavor (mis-)tagging efficiency            & $+1.7$                                & $-2.2$  \\
    \qquad Simulated sample size                      & $+4.2$                                & $-3.6$  \\
    \qquad Other experimental uncertainties           & $+2.1$                                & $-1.4$  \\
    [\cmsTabSkip]
    Theoretical                                       & $+3.9$                                & $-1.6$  \\
    \qquad \Vjets modeling                            & $+2.3$                                & $-1.2$  \\
    \qquad \PH modeling                               & $+3.2$                                & $-1.0$  \\
    [\cmsTabSkip]
    Total                                             & $+20.3$                               & $-19.9$ \\
  \end{scotch}
  \label{tab:systematics}
\end{table}

In conclusion, a search for standard model (SM) \PZ and Higgs bosons produced
with transverse momenta greater than 450\GeV and decaying to charm
quark-antiquark (\ccbar) pairs has been performed in a data sample corresponding
to an integrated luminosity of \analysisLumi at $\sqrt{s}=13\TeV$. New
algorithms based on deep neural networks have been developed to identify jets
originating from charm quark pairs. The \zcc process is observed in association
with jets at a hadron collider for the first time, with a signal strength of
\muZlineshort relative to the SM prediction. This observation establishes \zcc
as an important reference for future \xcc searches. An observed (expected)
upper limit on the product of the Higgs boson production cross section and
branching fraction to \ccbar of \muObsLimit (\muExpLimit) times the SM
expectation is set at 95\% confidence level.

\begin{acknowledgments}
We congratulate our colleagues in the CERN accelerator departments for the excellent performance of the LHC and thank the technical and administrative staffs at CERN and at other CMS institutes for their contributions to the success of the CMS effort. In addition, we gratefully acknowledge the computing centers and personnel of the Worldwide LHC Computing Grid and other centers for delivering so effectively the computing infrastructure essential to our analyses. Finally, we acknowledge the enduring support for the construction and operation of the LHC, the CMS detector, and the supporting computing infrastructure provided by the following funding agencies: BMBWF and FWF (Austria); FNRS and FWO (Belgium); CNPq, CAPES, FAPERJ, FAPERGS, and FAPESP (Brazil); MES and BNSF (Bulgaria); CERN; CAS, MoST, and NSFC (China); MINCIENCIAS (Colombia); MSES and CSF (Croatia); RIF (Cyprus); SENESCYT (Ecuador); MoER, ERC PUT and ERDF (Estonia); Academy of Finland, MEC, and HIP (Finland); CEA and CNRS/IN2P3 (France); BMBF, DFG, and HGF (Germany); GSRI (Greece); NKFIH (Hungary); DAE and DST (India); IPM (Iran); SFI (Ireland); INFN (Italy); MSIP and NRF (Republic of Korea); MES (Latvia); LAS (Lithuania); MOE and UM (Malaysia); BUAP, CINVESTAV, CONACYT, LNS, SEP, and UASLP-FAI (Mexico); MOS (Montenegro); MBIE (New Zealand); PAEC (Pakistan); MES and NSC (Poland); FCT (Portugal); MESTD (Serbia); MCIN/AEI and PCTI (Spain); MOSTR (Sri Lanka); Swiss Funding Agencies (Switzerland); MST (Taipei); MHESI and NSTDA (Thailand); TUBITAK and TENMAK (Turkey); NASU (Ukraine); STFC (United Kingdom); DOE and NSF (USA).
\end{acknowledgments}

\bibliography{auto_generated}

\providecommand{\href}[2]{#2}\begingroup\raggedright\begin{thebibliography}{10}%
\makeatletter
\providecommand{\hrefCMSnoop }[0]{\@secondoftwo}%
\makeatother
\providecommand{\doi}{\texttt{doi:}\begingroup \urlstyle{tt}\Url}

\bibitem{:2012gk}
\hrefCMSnoop {}{{ATLAS Collaboration}, ``Observation of a new particle in the
  search for the standard model {Higgs} boson with the {ATLAS} detector at the
  {LHC}'',} \textit{ Phys. Lett. B} \textbf{ 716} (2012) 1,
  \href{http://dx.doi.org/10.1016/j.physletb.2012.08.020}{\doi{10.1016/j.physletb.2012.08.020}},
  \href{http://www.arXiv.org/abs/1207.7214}{\texttt{arXiv:1207.7214}}.

\bibitem{:2012gu}
\hrefCMSnoop {}{{CMS Collaboration}, ``Observation of a new boson at a mass of
  {125\GeV} with the {CMS} experiment at the {LHC}'',} \textit{ Phys. Lett. B}
  \textbf{ 716} (2012) 30,
  \href{http://dx.doi.org/10.1016/j.physletb.2012.08.021}{\doi{10.1016/j.physletb.2012.08.021}},
  \href{http://www.arXiv.org/abs/1207.7235}{\texttt{arXiv:1207.7235}}.

\bibitem{Chatrchyan:2013lba}
\hrefCMSnoop {}{{CMS Collaboration}, ``{Observation of a new boson with mass
  near 125\GeV in $\Pp\Pp$ collisions at $\sqrt{s} = 7$ and 8\TeV}'',} \textit{
  JHEP} \textbf{ 06} (2013) 081,
  \href{http://dx.doi.org/10.1007/JHEP06(2013)081}{\doi{10.1007/JHEP06(2013)081}},
  \href{http://www.arXiv.org/abs/1303.4571}{\texttt{arXiv:1303.4571}}.

\bibitem{Sirunyan:2017khh}
\hrefCMSnoop {}{{CMS Collaboration}, ``{Observation of the Higgs boson decay to
  a pair of $\PGt$ leptons with the CMS detector}'',} \textit{ Phys. Lett. B}
  \textbf{ 779} (2018) 283,
  \href{http://dx.doi.org/10.1016/j.physletb.2018.02.004}{\doi{10.1016/j.physletb.2018.02.004}},
  \href{http://www.arXiv.org/abs/1708.00373}{\texttt{arXiv:1708.00373}}.

\bibitem{ATLAS:2022yrq}
\hrefCMSnoop {}{{ATLAS Collaboration}, ``{Measurements of Higgs boson
  production cross-sections in the $\PH\to\PGtp\PGtm$ decay channel in $\Pp\Pp$
  collisions at $\sqrt{s}=13\TeV$ with the ATLAS detector}'',} \textit{ JHEP}
  \textbf{ 08} (2022) 175,
  \href{http://dx.doi.org/10.1007/JHEP08(2022)175}{\doi{10.1007/JHEP08(2022)175}},
  \href{http://www.arXiv.org/abs/2201.08269}{\texttt{arXiv:2201.08269}}.

\bibitem{Sirunyan:2018hoz}
\hrefCMSnoop {}{{CMS Collaboration}, ``{Observation of $\ttbar\PH$
  Production}'',} \textit{ Phys. Rev. Lett.} \textbf{ 120} (2018) 231801,
  \href{http://dx.doi.org/10.1103/PhysRevLett.120.231801}{\doi{10.1103/PhysRevLett.120.231801}},
  \href{http://www.arXiv.org/abs/1804.02610}{\texttt{arXiv:1804.02610}}.

\bibitem{ATLAS:2018mme}
\hrefCMSnoop {}{{ATLAS Collaboration}, ``{Observation of Higgs boson production
  in association with a top quark pair at the LHC with the ATLAS detector}'',}
  \textit{ Phys. Lett. B} \textbf{ 784} (2018) 173,
  \href{http://dx.doi.org/10.1016/j.physletb.2018.07.035}{\doi{10.1016/j.physletb.2018.07.035}},
  \href{http://www.arXiv.org/abs/1806.00425}{\texttt{arXiv:1806.00425}}.

\bibitem{ATLAS:2018kot}
\hrefCMSnoop {}{{ATLAS Collaboration}, ``{Observation of $H \rightarrow
  b\bar{b}$ decays and $VH$ production with the ATLAS detector}'',} \textit{
  Phys. Lett. B} \textbf{ 786} (2018) 59,
  \href{http://dx.doi.org/10.1016/j.physletb.2018.09.013}{\doi{10.1016/j.physletb.2018.09.013}},
  \href{http://www.arXiv.org/abs/1808.08238}{\texttt{arXiv:1808.08238}}.

\bibitem{Sirunyan:2018kst}
\hrefCMSnoop {}{{CMS Collaboration}, ``{Observation} of {H}iggs {Boson} {Decay}
  to {Bottom} {Quarks}'',} \textit{ Phys. Rev. Lett.} \textbf{ 121} (2018)
  121801,
  \href{http://dx.doi.org/10.1103/PhysRevLett.121.121801}{\doi{10.1103/PhysRevLett.121.121801}},
  \href{http://www.arXiv.org/abs/1808.08242}{\texttt{arXiv:1808.08242}}.

\bibitem{Sirunyan:2018hbu}
\hrefCMSnoop {}{{CMS Collaboration}, ``{Search for the {Higgs} {Boson}
  {Decaying} to {Two} {Muons} in {Proton-Proton} {Collisions} at $\sqrt{s} =
  13\TeV$}'',} \textit{ Phys. Rev. Lett.} \textbf{ 122} (2019) 021801,
  \href{http://dx.doi.org/10.1103/PhysRevLett.122.021801}{\doi{10.1103/PhysRevLett.122.021801}},
  \href{http://www.arXiv.org/abs/1807.06325}{\texttt{arXiv:1807.06325}}.

\bibitem{2021135980}
\hrefCMSnoop {}{{ATLAS Collaboration}, ``A search for the dimuon decay of the
  standard model {Higgs} boson with the {ATLAS} detector'',} \textit{ {Phys.
  Lett. B}} \textbf{ {812}} ({2021}) {135980},
  \href{http://dx.doi.org/10.1016/j.physletb.2020.135980}{\doi{10.1016/j.physletb.2020.135980}},
  \href{http://www.arXiv.org/abs/2007.07830}{\texttt{arXiv:2007.07830}}.

\bibitem{Aaboud:2017ojs}
\hrefCMSnoop {}{{ATLAS Collaboration}, ``Search for the {Dimuon} {Decay} of the
  {Higgs} boson in {$\Pp\Pp$} {Collisions} at {$\sqrt{s} = 13\TeV$} with the
  {ATLAS} {Detector}'',} \textit{ Phys. Rev. Lett.} \textbf{ 119} (2017)
  051802,
  \href{http://dx.doi.org/10.1103/PhysRevLett.119.051802}{\doi{10.1103/PhysRevLett.119.051802}},
  \href{http://www.arXiv.org/abs/1705.04582}{\texttt{arXiv:1705.04582}}.

\bibitem{cmscollaboration2020evidence}
\hrefCMSnoop {}{{CMS Collaboration}, ``{Evidence for Higgs boson decay to a
  pair of muons}'',} \textit{ JHEP} \textbf{ 01} (2021) 148,
  \href{http://dx.doi.org/10.1007/JHEP01(2021)148}{\doi{10.1007/JHEP01(2021)148}},
  \href{http://www.arXiv.org/abs/2009.04363}{\texttt{arXiv:2009.04363}}.

\bibitem{CMS-DP-2018-046}
\href {https://cds.cern.ch/record/2630438}{{CMS Collaboration}, ``Performance
  of deep tagging algorithms for boosted double quark jet topology in
  proton-proton collisions at {13\TeV} with the {Phase-0} {CMS} detector'',}
  CMS Detector Performance Note CMS-DP-2018-046, 2018.

\bibitem{Aaboud_2018}
\hrefCMSnoop {}{{ATLAS Collaboration}, ``Search for the {Decay} of the {Higgs}
  {Boson} to {Charm} {Quarks} with the {ATLAS} {Experiment}'',} \textit{ Phys.
  Rev. Lett.} \textbf{ 120} (2018) 211802,
  \href{http://dx.doi.org/10.1103/PhysRevLett.120.211802}{\doi{10.1103/PhysRevLett.120.211802}},
  \href{http://www.arXiv.org/abs/1802.04329}{\texttt{arXiv:1802.04329}}.

\bibitem{ATLAS:2022ers}
\hrefCMSnoop {}{{ATLAS Collaboration}, ``{Direct constraint on the Higgs-charm
  coupling from a search for Higgs boson decays into charm quarks with the
  ATLAS detector}'',} \textit{ Eur. Phys. J. C} \textbf{ 82} (2022) 717,
  \href{http://dx.doi.org/10.1140/epjc/s10052-022-10588-3}{\doi{10.1140/epjc/s10052-022-10588-3}},
  \href{http://www.arXiv.org/abs/2201.11428}{\texttt{arXiv:2201.11428}}.

\bibitem{Sirunyan:2019qia}
\hrefCMSnoop {}{{CMS Collaboration}, ``{A search for the standard model Higgs
  boson decaying to charm quarks}'',} \textit{ JHEP} \textbf{ 03} (2020) 131,
  \href{http://dx.doi.org/10.1007/JHEP03(2020)131}{\doi{10.1007/JHEP03(2020)131}},
  \href{http://www.arXiv.org/abs/1912.01662}{\texttt{arXiv:1912.01662}}.

\bibitem{CMS:2022psv}
\hrefCMSnoop {}{{CMS Collaboration}, ``{Search for Higgs boson decay to a charm
  quark-antiquark pair in proton-proton collisions at $\sqrt{s} = 13\TeV$}'',}
  2022. \href{http://www.arXiv.org/abs/2205.05550}{\texttt{arXiv:2205.05550}}.
  \textit{Phys. Rev. Lett.} (to be published).

\bibitem{Sirunyan:2017dgc}
\hrefCMSnoop {}{{CMS Collaboration}, ``{Inclusive} {Search} for a {Highly}
  {Boosted} {Higgs} {Boson} {Decaying} to a {Bottom} {Quark-Antiquark}
  {Pair}'',} \textit{ Phys. Rev. Lett.} \textbf{ 120} (2018) 071802,
  \href{http://dx.doi.org/10.1103/PhysRevLett.120.071802}{\doi{10.1103/PhysRevLett.120.071802}},
  \href{http://www.arXiv.org/abs/1709.05543}{\texttt{arXiv:1709.05543}}.

\bibitem{CMS:2020zge}
\hrefCMSnoop {}{{CMS Collaboration}, ``{Inclusive search for highly boosted
  Higgs bosons decaying to bottom quark-antiquark pairs in proton-proton
  collisions at $\sqrt{s} = 13\TeV$}'',} \textit{ JHEP} \textbf{ 12} (2020)
  085,
  \href{http://dx.doi.org/10.1007/JHEP12(2020)085}{\doi{10.1007/JHEP12(2020)085}},
  \href{http://www.arXiv.org/abs/2006.13251}{\texttt{arXiv:2006.13251}}.

\bibitem{Chatrchyan:2008zzk}
\hrefCMSnoop {}{{CMS Collaboration}, ``The {CMS} experiment at the {CERN}
  {LHC}'',} \textit{ JINST} \textbf{ 3} (2008) S08004,
  \href{http://dx.doi.org/10.1088/1748-0221/3/08/S08004}{\doi{10.1088/1748-0221/3/08/S08004}}.

\bibitem{Khachatryan:2016bia}
\hrefCMSnoop {}{{CMS Collaboration}, ``The {CMS} trigger system'',} \textit{
  JINST} \textbf{ 12} (2017) P01020,
  \href{http://dx.doi.org/10.1088/1748-0221/12/01/P01020}{\doi{10.1088/1748-0221/12/01/P01020}},
  \href{http://www.arXiv.org/abs/1609.02366}{\texttt{arXiv:1609.02366}}.

\bibitem{CMS:2020cmk}
\hrefCMSnoop {}{{CMS Collaboration}, ``{Performance of the CMS Level-1 trigger
  in proton-proton collisions at $\sqrt{s} = 13\TeV$}'',} \textit{ JINST}
  \textbf{ 15} (2020) P10017,
  \href{http://dx.doi.org/10.1088/1748-0221/15/10/P10017}{\doi{10.1088/1748-0221/15/10/P10017}},
  \href{http://www.arXiv.org/abs/2006.10165}{\texttt{arXiv:2006.10165}}.

\bibitem{CMS:2020uim}
\hrefCMSnoop {}{{CMS Collaboration}, ``Electron and photon reconstruction and
  identification with the {CMS} experiment at the {CERN} {LHC}'',} \textit{
  JINST} \textbf{ 16} (2021) P05014,
  \href{http://dx.doi.org/10.1088/1748-0221/16/05/P05014}{\doi{10.1088/1748-0221/16/05/P05014}},
  \href{http://www.arXiv.org/abs/2012.06888}{\texttt{arXiv:2012.06888}}.

\bibitem{Sirunyan:2018fpa}
\hrefCMSnoop {}{{CMS Collaboration}, ``{Performance of the CMS muon detector
  and muon reconstruction with proton-proton collisions at $\sqrt{s}=
  13\TeV$}'',} \textit{ JINST} \textbf{ 13} (2018) P06015,
  \href{http://dx.doi.org/10.1088/1748-0221/13/06/P06015}{\doi{10.1088/1748-0221/13/06/P06015}},
  \href{http://www.arXiv.org/abs/1804.04528}{\texttt{arXiv:1804.04528}}.

\bibitem{Chatrchyan:2014fea}
\hrefCMSnoop {}{{CMS Collaboration}, ``{Description and performance of track
  and primary-vertex reconstruction with the CMS tracker}'',} \textit{ JINST}
  \textbf{ 9} (2014) P10009,
  \href{http://dx.doi.org/10.1088/1748-0221/9/10/P10009}{\doi{10.1088/1748-0221/9/10/P10009}},
  \href{http://www.arXiv.org/abs/1405.6569}{\texttt{arXiv:1405.6569}}.

\bibitem{Sirunyan:2017ulk}
\hrefCMSnoop {}{{CMS Collaboration}, ``Particle-flow reconstruction and global
  event description with the {CMS} detector'',} \textit{ JINST} \textbf{ 12}
  (2017) P10003,
  \href{http://dx.doi.org/10.1088/1748-0221/12/10/P10003}{\doi{10.1088/1748-0221/12/10/P10003}},
  \href{http://www.arXiv.org/abs/1706.04965}{\texttt{arXiv:1706.04965}}.

\bibitem{CMS:2018jrd}
\hrefCMSnoop {}{{CMS Collaboration}, ``{Performance of reconstruction and
  identification of $\PGt$ leptons decaying to hadrons and $\PGnGt$ in $\Pp\Pp$
  collisions at $\sqrt{s}= 13\TeV$}'',} \textit{ JINST} \textbf{ 13} (2018)
  P10005,
  \href{http://dx.doi.org/10.1088/1748-0221/13/10/P10005}{\doi{10.1088/1748-0221/13/10/P10005}},
  \href{http://www.arXiv.org/abs/1809.02816}{\texttt{arXiv:1809.02816}}.

\bibitem{Khachatryan:2016kdb}
\hrefCMSnoop {}{{CMS Collaboration}, ``{Jet energy scale and resolution in the
  CMS experiment in $\Pp\Pp$ collisions at 8\TeV}'',} \textit{ JINST} \textbf{
  12} (2017) P02014,
  \href{http://dx.doi.org/10.1088/1748-0221/12/02/P02014}{\doi{10.1088/1748-0221/12/02/P02014}},
  \href{http://www.arXiv.org/abs/1607.03663}{\texttt{arXiv:1607.03663}}.

\bibitem{CMS:2019ctu}
\hrefCMSnoop {}{{CMS Collaboration}, ``Performance of missing transverse
  momentum reconstruction in proton-proton collisions at {$\sqrt{s} = 13\TeV$}
  using the {CMS} detector'',} \textit{ JINST} \textbf{ 14} (2019) P07004,
  \href{http://dx.doi.org/10.1088/1748-0221/14/07/P07004}{\doi{10.1088/1748-0221/14/07/P07004}},
  \href{http://www.arXiv.org/abs/1903.06078}{\texttt{arXiv:1903.06078}}.

\bibitem{Alwall:2014hca}
J.~Alwall\hrefCMSnoop {}{ {et~al.}, ``{The automated computation of tree-level
  and next-to-leading order differential cross sections, and their matching to
  parton shower simulations}'',} \textit{ JHEP} \textbf{ 07} (2014) 079,
  \href{http://dx.doi.org/10.1007/JHEP07(2014)079}{\doi{10.1007/JHEP07(2014)079}},
  \href{http://www.arXiv.org/abs/1405.0301}{\texttt{arXiv:1405.0301}}.

\bibitem{Alwall:2007fs}
J.~Alwall\hrefCMSnoop {}{ {et~al.}, ``Comparative study of various algorithms
  for the merging of parton showers and matrix elements in hadronic
  collisions'',} \textit{ Eur. Phys. J. C} \textbf{ 53} (2008) 473,
  \href{http://dx.doi.org/10.1140/epjc/s10052-007-0490-5}{\doi{10.1140/epjc/s10052-007-0490-5}},
  \href{http://www.arXiv.org/abs/0706.2569}{\texttt{arXiv:0706.2569}}.

\bibitem{Nason:2004rx}
\hrefCMSnoop {}{P.~Nason, ``A new method for combining {NLO QCD} with shower
  {Monte Carlo} algorithms'',} \textit{ JHEP} \textbf{ 11} (2004) 040,
  \href{http://dx.doi.org/10.1088/1126-6708/2004/11/040}{\doi{10.1088/1126-6708/2004/11/040}},
  \href{http://www.arXiv.org/abs/hep-ph/0409146}{\texttt{arXiv:hep-ph/0409146}}.

\bibitem{Frixione:2007nw}
\hrefCMSnoop {}{S.~Frixione, P.~Nason, and G.~Ridolfi, ``A positive-weight
  next-to-leading-order {Monte Carlo} for heavy flavour hadroproduction'',}
  \textit{ JHEP} \textbf{ 09} (2007) 126,
  \href{http://dx.doi.org/10.1088/1126-6708/2007/09/126}{\doi{10.1088/1126-6708/2007/09/126}},
  \href{http://www.arXiv.org/abs/0707.3088}{\texttt{arXiv:0707.3088}}.

\bibitem{Frixione:2007vw}
\hrefCMSnoop {}{S.~Frixione, P.~Nason, and C.~Oleari, ``Matching {NLO} {QCD}
  computations with parton shower simulations: {The} {\POWHEG} method'',}
  \textit{ JHEP} \textbf{ 11} (2007) 070,
  \href{http://dx.doi.org/10.1088/1126-6708/2007/11/070}{\doi{10.1088/1126-6708/2007/11/070}},
  \href{http://www.arXiv.org/abs/0709.2092}{\texttt{arXiv:0709.2092}}.

\bibitem{Alioli:2010xd}
\hrefCMSnoop {}{S.~Alioli, P.~Nason, C.~Oleari, and E.~Re, ``A general
  framework for implementing {NLO} calculations in shower {Monte Carlo}
  programs: the {\POWHEG} {BOX}'',} \textit{ JHEP} \textbf{ 06} (2010) 043,
  \href{http://dx.doi.org/10.1007/JHEP06(2010)043}{\doi{10.1007/JHEP06(2010)043}},
  \href{http://www.arXiv.org/abs/1002.2581}{\texttt{arXiv:1002.2581}}.

\bibitem{Re:2010bp}
\hrefCMSnoop {}{E.~Re, ``Single-top {$\PW\PQt$}-channel production matched with
  parton showers using the {\POWHEG} method'',} \textit{ Eur. Phys. J. C}
  \textbf{ 71} (2011) 1547,
  \href{http://dx.doi.org/10.1140/epjc/s10052-011-1547-z}{\doi{10.1140/epjc/s10052-011-1547-z}},
  \href{http://www.arXiv.org/abs/1009.2450}{\texttt{arXiv:1009.2450}}.

\bibitem{Frederix:2012dh}
\hrefCMSnoop {}{R.~Frederix, E.~Re, and P.~Torrielli, ``Single-top $t$-channel
  hadroproduction in the four-flavour scheme with {\POWHEG} and {a\MCATNLO}'',}
  \textit{ JHEP} \textbf{ 09} (2012) 130,
  \href{http://dx.doi.org/10.1007/JHEP09(2012)130}{\doi{10.1007/JHEP09(2012)130}},
  \href{http://www.arXiv.org/abs/1207.5391}{\texttt{arXiv:1207.5391}}.

\bibitem{Sjostrand:2014zea}
T.~Sj{\"o}strand\hrefCMSnoop {}{ {et~al.}, ``An introduction to
  {\PYTHIA8.2}'',} \textit{ Comput. Phys. Commun.} \textbf{ 191} (2015) 159,
  \href{http://dx.doi.org/10.1016/j.cpc.2015.01.024}{\doi{10.1016/j.cpc.2015.01.024}},
  \href{http://www.arXiv.org/abs/1410.3012}{\texttt{arXiv:1410.3012}}.

\bibitem{Frederix:2012ps}
\hrefCMSnoop {}{R.~Frederix and S.~Frixione, ``{Merging meets matching in
  MC@NLO}'',} \textit{ JHEP} \textbf{ 12} (2012) 061,
  \href{http://dx.doi.org/10.1007/JHEP12(2012)061}{\doi{10.1007/JHEP12(2012)061}},
  \href{http://www.arXiv.org/abs/1209.6215}{\texttt{arXiv:1209.6215}}.

\bibitem{Kallweit:2014xda}
S.~Kallweit\hrefCMSnoop {}{ {et~al.}, ``{NLO electroweak automation and precise
  predictions for $\PW$+multijet production at the LHC}'',} \textit{ JHEP}
  \textbf{ 04} (2015) 012,
  \href{http://dx.doi.org/10.1007/JHEP04(2015)012}{\doi{10.1007/JHEP04(2015)012}},
  \href{http://www.arXiv.org/abs/1412.5157}{\texttt{arXiv:1412.5157}}.

\bibitem{Kallweit:2015fta}
S.~Kallweit\href
  {https://inspirehep.net/record/1372103/files/arXiv:1505.05704.pdf}{ {et~al.},
  ``{NLO QCD+EW automation and precise predictions for $\PV$+multijet
  production}'',} in \textit{ 50th {Rencontres} de {Moriond}, {QCD} and high
  energy interactions}, p.~121.
\newblock 2015.
\newblock
  \href{http://www.arXiv.org/abs/1505.05704}{\texttt{arXiv:1505.05704}}.

\bibitem{Kallweit:2015dum}
S.~Kallweit\hrefCMSnoop {}{ {et~al.}, ``{NLO QCD+EW predictions for $\PV$+jets
  including off-shell vector-boson decays and multijet merging}'',} \textit{
  JHEP} \textbf{ 04} (2016) 021,
  \href{http://dx.doi.org/10.1007/JHEP04(2016)021}{\doi{10.1007/JHEP04(2016)021}},
  \href{http://www.arXiv.org/abs/1511.08692}{\texttt{arXiv:1511.08692}}.

\bibitem{Lindert:2017olm}
\hrefCMSnoop {}{J.~M. Lindert {et~al.}, ``{Precise predictions for $\PV$+jets
  dark matter backgrounds}'',} \textit{ Eur. Phys. J. C} \textbf{ 77} (2017)
  829,
  \href{http://dx.doi.org/10.1140/epjc/s10052-017-5389-1}{\doi{10.1140/epjc/s10052-017-5389-1}},
  \href{http://www.arXiv.org/abs/1705.04664}{\texttt{arXiv:1705.04664}}.

\bibitem{Campbell:2010ff}
\hrefCMSnoop {}{J.~M. Campbell and R.~K. Ellis, ``{\MCFM} for the {Tevatron}
  and the {LHC}'',} \textit{ Nucl. Phys. Proc. Suppl.} \textbf{ 205-206} (2010)
  10,
  \href{http://dx.doi.org/10.1016/j.nuclphysbps.2010.08.011}{\doi{10.1016/j.nuclphysbps.2010.08.011}},
  \href{http://www.arXiv.org/abs/1007.3492}{\texttt{arXiv:1007.3492}}.

\bibitem{Hamilton:2012rf}
\hrefCMSnoop {}{K.~Hamilton, P.~Nason, C.~Oleari, and G.~Zanderighi, ``{Merging
  \PH/\PW/\PZ + 0 and 1 jet at NLO with no merging scale: a path to parton
  shower + NNLO matching}'',} \textit{ JHEP} \textbf{ 05} (2013) 082,
  \href{http://dx.doi.org/10.1007/JHEP05(2013)082}{\doi{10.1007/JHEP05(2013)082}},
  \href{http://www.arXiv.org/abs/1212.4504}{\texttt{arXiv:1212.4504}}.

\bibitem{Luisoni:2013kna}
\hrefCMSnoop {}{G.~Luisoni, P.~Nason, C.~Oleari, and F.~Tramontano,
  ``{$\PH\PWpm$/$\PH\PZ$}+0 and 1 jet at {NLO} with the {\POWHEG box}
  interfaced to {GoSam} and their merging within {\MINLO}'',} \textit{ JHEP}
  \textbf{ 10} (2013) 083,
  \href{http://dx.doi.org/10.1007/JHEP10(2013)083}{\doi{10.1007/JHEP10(2013)083}},
  \href{http://www.arXiv.org/abs/1306.2542}{\texttt{arXiv:1306.2542}}.

\bibitem{Becker:2669113}
\hrefCMSnoop {}{K.~Becker {et~al.}, ``{Precise predictions for boosted Higgs
  production}'',} 2020.
  \href{http://www.arXiv.org/abs/2005.07762}{\texttt{arXiv:2005.07762}}.

\bibitem{Bagnaschi2012}
\hrefCMSnoop {}{E.~Bagnaschi, G.~Degrassi, P.~Slavich, and A.~Vicini, ``{Higgs}
  production via gluon fusion in the {\POWHEG} approach in the {SM} and in the
  {MSSM}'',} \textit{ JHEP} \textbf{ 02} (2012) 088,
  \href{http://dx.doi.org/10.1007/JHEP02(2012)088}{\doi{10.1007/JHEP02(2012)088}},
  \href{http://www.arXiv.org/abs/1111.2854}{\texttt{arXiv:1111.2854}}.

\bibitem{Nason:2009ai}
\hrefCMSnoop {}{P.~Nason and C.~Oleari, ``{NLO} {Higgs} boson production via
  vector-boson fusion matched with shower in {\POWHEG}'',} \textit{ JHEP}
  \textbf{ 02} (2010) 037,
  \href{http://dx.doi.org/10.1007/JHEP02(2010)037}{\doi{10.1007/JHEP02(2010)037}},
  \href{http://www.arXiv.org/abs/0911.5299}{\texttt{arXiv:0911.5299}}.

\bibitem{Hartanto:2015uka}
\hrefCMSnoop {}{H.~B. Hartanto, B.~J{\"a}ger, L.~Reina, and D.~Wackeroth,
  ``{Higgs} boson production in association with top quarks in the {\POWHEG}
  {BOX}'',} \textit{ Phys. Rev. D} \textbf{ 91} (2015) 094003,
  \href{http://dx.doi.org/10.1103/PhysRevD.91.094003}{\doi{10.1103/PhysRevD.91.094003}},
  \href{http://www.arXiv.org/abs/1501.04498}{\texttt{arXiv:1501.04498}}.

\bibitem{Khachatryan:2015pea}
\hrefCMSnoop {}{{CMS Collaboration}, ``{Event generator tunes obtained from
  underlying event and multiparton scattering measurements}'',} \textit{ Eur.
  Phys. J. C} \textbf{ 76} (2016) 155,
  \href{http://dx.doi.org/10.1140/epjc/s10052-016-3988-x}{\doi{10.1140/epjc/s10052-016-3988-x}},
  \href{http://www.arXiv.org/abs/1512.00815}{\texttt{arXiv:1512.00815}}.

\bibitem{Sirunyan:2019dfx}
\hrefCMSnoop {}{{CMS Collaboration}, ``Extraction and validation of a new set
  of {CMS} {\PYTHIA}8 tunes from underlying-event measurements'',} \textit{
  Eur. Phys. J. C} \textbf{ 80} (2020) 4,
  \href{http://dx.doi.org/10.1140/epjc/s10052-019-7499-4}{\doi{10.1140/epjc/s10052-019-7499-4}},
  \href{http://www.arXiv.org/abs/1903.12179}{\texttt{arXiv:1903.12179}}.

\bibitem{Ball:2014uwa}
\hrefCMSnoop {}{{NNPDF} Collaboration, ``{Parton distributions for the LHC run
  II}'',} \textit{ JHEP} \textbf{ 04} (2015) 040,
  \href{http://dx.doi.org/10.1007/JHEP04(2015)040}{\doi{10.1007/JHEP04(2015)040}},
  \href{http://www.arXiv.org/abs/1410.8849}{\texttt{arXiv:1410.8849}}.

\bibitem{Ball:2017nwa}
\hrefCMSnoop {}{{NNPDF} Collaboration, ``{Parton distributions from
  high-precision collider data}'',} \textit{ Eur. Phys. J. C} \textbf{ 77}
  (2017) 663,
  \href{http://dx.doi.org/10.1140/epjc/s10052-017-5199-5}{\doi{10.1140/epjc/s10052-017-5199-5}},
  \href{http://www.arXiv.org/abs/1706.00428}{\texttt{arXiv:1706.00428}}.

\bibitem{GEANT4}
\hrefCMSnoop {}{{GEANT4} Collaboration, ``{\GEANTfour}---a simulation
  toolkit'',} \textit{ Nucl. Instrum. Methods Phys. Res. Sect. A} \textbf{ 506}
  (2003) 250,
  \href{http://dx.doi.org/10.1016/S0168-9002(03)01368-8}{\doi{10.1016/S0168-9002(03)01368-8}}.

\bibitem{Cacciari:2008gp}
\hrefCMSnoop {}{M.~Cacciari, G.~P. Salam, and G.~Soyez, ``The anti-\kt jet
  clustering algorithm'',} \textit{ JHEP} \textbf{ 04} (2008) 063,
  \href{http://dx.doi.org/10.1088/1126-6708/2008/04/063}{\doi{10.1088/1126-6708/2008/04/063}},
  \href{http://www.arXiv.org/abs/0802.1189}{\texttt{arXiv:0802.1189}}.

\bibitem{Cacciari:2011ma}
\hrefCMSnoop {}{M.~Cacciari, G.~P. Salam, and G.~Soyez, ``{FastJet} user
  manual'',} \textit{ Eur. Phys. J. C} \textbf{ 72} (2012) 1896,
  \href{http://dx.doi.org/10.1140/epjc/s10052-012-1896-2}{\doi{10.1140/epjc/s10052-012-1896-2}},
  \href{http://www.arXiv.org/abs/1111.6097}{\texttt{arXiv:1111.6097}}.

\bibitem{Bertolini:2014bba}
\hrefCMSnoop {}{D.~Bertolini, P.~Harris, M.~Low, and N.~Tran, ``{Pileup per
  particle identification}'',} \textit{ JHEP} \textbf{ 10} (2014) 059,
  \href{http://dx.doi.org/10.1007/JHEP10(2014)059}{\doi{10.1007/JHEP10(2014)059}},
  \href{http://www.arXiv.org/abs/1407.6013}{\texttt{arXiv:1407.6013}}.

\bibitem{Larkoski:2014wba}
\hrefCMSnoop {}{A.~J. Larkoski, S.~Marzani, G.~Soyez, and J.~Thaler, ``{Soft
  drop}'',} \textit{ JHEP} \textbf{ 05} (2014) 146,
  \href{http://dx.doi.org/10.1007/JHEP05(2014)146}{\doi{10.1007/JHEP05(2014)146}},
  \href{http://www.arXiv.org/abs/1402.2657}{\texttt{arXiv:1402.2657}}.

\bibitem{CMS:2019ied}
\hrefCMSnoop {}{{CMS Collaboration}, ``{Performance of the reconstruction and
  identification of high-momentum muons in proton-proton collisions at
  $\sqrt{s} = 13\TeV$}'',} \textit{ JINST} \textbf{ 15} (2020) P02027,
  \href{http://dx.doi.org/10.1088/1748-0221/15/02/P02027}{\doi{10.1088/1748-0221/15/02/P02027}},
  \href{http://www.arXiv.org/abs/1912.03516}{\texttt{arXiv:1912.03516}}.

\bibitem{Sirunyan:2017ezt}
\hrefCMSnoop {}{{CMS Collaboration}, ``Identification of heavy-flavour jets
  with the {CMS} detector in {$\Pp\Pp$} collisions at {13\TeV}'',} \textit{
  JINST} \textbf{ 13} (2018) P05011,
  \href{http://dx.doi.org/10.1088/1748-0221/13/05/P05011}{\doi{10.1088/1748-0221/13/05/P05011}},
  \href{http://www.arXiv.org/abs/1712.07158}{\texttt{arXiv:1712.07158}}.

\bibitem{Dasgupta:2013ihk}
\hrefCMSnoop {}{M.~Dasgupta, A.~Fregoso, S.~Marzani, and G.~P. Salam,
  ``{Towards an understanding of jet substructure}'',} \textit{ JHEP} \textbf{
  09} (2013) 029,
  \href{http://dx.doi.org/10.1007/JHEP09(2013)029}{\doi{10.1007/JHEP09(2013)029}},
  \href{http://www.arXiv.org/abs/1307.0007}{\texttt{arXiv:1307.0007}}.

\bibitem{Dolen:2016kst}
J.~Dolen\hrefCMSnoop {}{ {et~al.}, ``Thinking outside the {ROC}s: Designing
  decorrelated taggers ({DDT}) for jet substructure'',} \textit{ JHEP} \textbf{
  05} (2016) 156,
  \href{http://dx.doi.org/10.1007/JHEP05(2016)156}{\doi{10.1007/JHEP05(2016)156}},
  \href{http://www.arXiv.org/abs/1603.00027}{\texttt{arXiv:1603.00027}}.

\bibitem{ATLAS:2019mgf}
\hrefCMSnoop {}{{ATLAS Collaboration}, ``{Measurement of soft-drop jet
  observables in $\Pp\Pp$ collisions with the ATLAS detector at $\sqrt{s}
  =13\TeV$}'',} \textit{ Phys. Rev. D} \textbf{ 101} (2020) 052007,
  \href{http://dx.doi.org/10.1103/PhysRevD.101.052007}{\doi{10.1103/PhysRevD.101.052007}},
  \href{http://www.arXiv.org/abs/1912.09837}{\texttt{arXiv:1912.09837}}.

\bibitem{Moult:2016cvt}
\hrefCMSnoop {}{I.~Moult, L.~Necib, and J.~Thaler, ``{New angles on energy
  correlation functions}'',} \textit{ JHEP} \textbf{ 12} (2016) 153,
  \href{http://dx.doi.org/10.1007/JHEP12(2016)153}{\doi{10.1007/JHEP12(2016)153}},
  \href{http://www.arXiv.org/abs/1609.07483}{\texttt{arXiv:1609.07483}}.

\bibitem{Larkoski:2013eya}
\hrefCMSnoop {}{A.~J. Larkoski, G.~P. Salam, and J.~Thaler, ``{Energy
  correlation functions for jet substructure}'',} \textit{ JHEP} \textbf{ 06}
  (2013) 108,
  \href{http://dx.doi.org/10.1007/JHEP06(2013)108}{\doi{10.1007/JHEP06(2013)108}},
  \href{http://www.arXiv.org/abs/1305.0007}{\texttt{arXiv:1305.0007}}.

\bibitem{CMS-DP-2022-041}
\href {https://cds.cern.ch/record/2839736}{{CMS Collaboration}, ``Performance
  of the mass-decorrelated {DeepDoubleX} classifier for double-{\PQb} and
  double-{\PQc} large-radius jets with the {CMS} detector'',} CMS Detector
  Performance Note CMS-DP-2022-041, 2022.

\bibitem{integrated}
\href {http://proceedings.mlr.press/v70/sundararajan17a.html}{M.~Sundararajan,
  A.~Taly, and Q.~Yan, ``Axiomatic attribution for deep networks'',} in
  \textit{ Proceedings of the 34th International Conference on Machine
  Learning}, p.~3319.
\newblock PMLR, 2017.
\newblock
  \href{http://www.arXiv.org/abs/1703.01365}{\texttt{arXiv:1703.01365}}.

\bibitem{MONTAVON2017211}
G.~Montavon\hrefCMSnoop {}{ {et~al.}, ``{Explaining nonlinear classification
  decisions with deep Taylor decomposition}'',} \textit{ Pattern Recognit.}
  \textbf{ 65} (2017) 211,
  \href{http://dx.doi.org/10.1016/j.patcog.2016.11.008}{\doi{10.1016/j.patcog.2016.11.008}},
  \href{http://www.arXiv.org/abs/1512.02479}{\texttt{arXiv:1512.02479}}.

\bibitem{Thaler:2010tr}
\hrefCMSnoop {}{J.~Thaler and K.~Van~Tilburg, ``{Identifying boosted objects
  with N-subjettiness}'',} \textit{ JHEP} \textbf{ 03} (2011) 015,
  \href{http://dx.doi.org/10.1007/JHEP03(2011)015}{\doi{10.1007/JHEP03(2011)015}},
  \href{http://www.arXiv.org/abs/1011.2268}{\texttt{arXiv:1011.2268}}.

\bibitem{bernstein1912}
\hrefCMSnoop {}{S.~Bernstein, ``D{\'{e}}monstration du th{\'{e}}or{\`{e}}me de
  {Weierstrass} fond{\'{e}}e sur le calcul des probabilit{\'{e}}s'',} \textit{
  Comm. Kharkov Math. Soc.} \textbf{ 13} (1912) 01.

\bibitem{ref:ftest}
\hrefCMSnoop {}{R.~A. Fisher, ``On the interpretation of $\chi^{2}$ from
  contingency tables, and the calculation of {P}'',} \textit{ J. R. Stat. Soc.}
  \textbf{ 85} (1922) 87,
  \href{http://dx.doi.org/10.2307/2340521}{\doi{10.2307/2340521}}.

\bibitem{CMS:jecs}
\href {https://cds.cern.ch/record/2715872}{{CMS Collaboration}, ``{Jet energy
  scale and resolution performance with 13\TeV data collected by CMS in
  2016--2018}'',} CMS Detector Performance Note CMS-DP-2020-019, 2020.

\bibitem{CMS:btv16}
\href {http://cds.cern.ch/record/2263801}{{CMS Collaboration}, ``{Performance
  of heavy flavour identification algorithms in proton-proton collisions at
  13\TeV at the CMS experiment}'',} CMS Detector Performance Note
  CMS-DP-2017-012, 2017.

\bibitem{CMS:btv1718}
\href {https://cds.cern.ch/record/2627468}{{CMS Collaboration}, ``{Performance
  of \PQb-tagging algorithms in proton-proton collisions at 13\TeV with Phase 1
  CMS detector}'',} CMS Detector Performance Note CMS-DP-2018-033, 2018.

\bibitem{CMS:pileup}
\href {https://cds.cern.ch/record/2715906}{{CMS Collaboration}, ``{Performance
  of the pile up jet identification in CMS for Run 2}'',} CMS Detector
  Performance Note CMS-DP-2020-020, 2020.

\bibitem{Barlow:1993dm}
\hrefCMSnoop {}{R.~J. Barlow and C.~Beeston, ``{Fitting using finite Monte
  Carlo samples}'',} \textit{ Comput. Phys. Commun.} \textbf{ 77} (1993) 219,
  \href{http://dx.doi.org/10.1016/0010-4655(93)90005-W}{\doi{10.1016/0010-4655(93)90005-W}}.

\bibitem{CMS:lumi}
\hrefCMSnoop {}{{CMS Collaboration}, ``{Precision luminosity measurement in
  proton-proton collisions at $\sqrt{s} = 13\TeV$ in 2015 and 2016 at CMS}'',}
  \textit{ Eur. Phys. J. C} \textbf{ 81} (2021) 800,
  \href{http://dx.doi.org/10.1140/epjc/s10052-021-09538-2}{\doi{10.1140/epjc/s10052-021-09538-2}},
  \href{http://www.arXiv.org/abs/2104.01927}{\texttt{arXiv:2104.01927}}.

\bibitem{CMS:lumi17}
\href {http://cds.cern.ch/record/2621960}{{CMS Collaboration}, ``{CMS
  luminosity measurement for the 2017 data-taking period at $\sqrt{s} =
  13\TeV$}'',} CMS Physics Analysis Summary CMS-PAS-LUM-17-004, 2018.

\bibitem{CMS:lumi18}
\href {http://cds.cern.ch/record/2676164}{{CMS Collaboration}, ``{CMS
  luminosity measurement for the 2018 data-taking period at $\sqrt{s} =
  13\TeV$}'',} CMS Physics Analysis Summary CMS-PAS-LUM-18-002, 2019.

\bibitem{LHCCLs}
\href {https://cds.cern.ch/record/1379837}{{The ATLAS Collaboration, The CMS
  Collaboration, The LHC Higgs Combination Group}, ``Procedure for the {LHC}
  {Higgs} boson search combination in {Summer} 2011'',} Technical Report
  CMS-NOTE-2011-005, ATL-PHYS-PUB-2011-11, 2011.

\bibitem{CMS:2020hkd}
\hrefCMSnoop {}{{CMS Collaboration}, ``{Measurement of the \PZ boson
  differential production cross section using its invisible decay mode
  ($\PZ\to\PGn\PAGn$) in proton-proton collisions at $\sqrt{s}= 13\TeV$}'',}
  \textit{ JHEP} \textbf{ 05} (2021) 205,
  \href{http://dx.doi.org/10.1007/JHEP05(2021)205}{\doi{10.1007/JHEP05(2021)205}},
  \href{http://www.arXiv.org/abs/2012.09254}{\texttt{arXiv:2012.09254}}.

\bibitem{Workman:2022ynf}
\hrefCMSnoop {}{{{Particle Data Group}}, ``{Review} of particle physics'',}
  \textit{ Prog. Theor. Exp. Phys.} 083C01,
  \href{http://dx.doi.org/10.1093/ptep/ptac097}{\doi{10.1093/ptep/ptac097}}.

\bibitem{CLS2}
\hrefCMSnoop {}{T.~Junk, ``{Confidence level computation for combining searches
  with small statistics}'',} \textit{ Nucl. Instrum. Methods Phys. Res. Sect.
  A} \textbf{ 434} (1999) 435,
  \href{http://dx.doi.org/10.1016/S0168-9002(99)00498-2}{\doi{10.1016/S0168-9002(99)00498-2}},
  \href{http://www.arXiv.org/abs/hep-ex/9902006}{\texttt{arXiv:hep-ex/9902006}}.

\bibitem{CLS1}
\hrefCMSnoop {}{A.~L. Read, ``{Presentation of search results: the \CLs
  technique}'',} \textit{ J. Phys. G} \textbf{ 28} (2002) 2693,
  \href{http://dx.doi.org/10.1088/0954-3899/28/10/313}{\doi{10.1088/0954-3899/28/10/313}}.

\bibitem{Cowan:2010js}
\hrefCMSnoop {}{G.~Cowan, K.~Cranmer, E.~Gross, and O.~Vitells, ``Asymptotic
  formulae for likelihood-based tests of new physics'',} \textit{ Eur. Phys. J.
  C} \textbf{ 71} (2011) 1554,
  \href{http://dx.doi.org/10.1140/epjc/s10052-011-1554-0}{\doi{10.1140/epjc/s10052-011-1554-0}},
  \href{http://www.arXiv.org/abs/1007.1727}{\texttt{arXiv:1007.1727}}.
  [Erratum: \DOI{10.1140/epjc/s10052-013-2501-z}].

\bibitem{hepdata}
\hrefCMSnoop {}{``{HEPD}ata record for this analysis'',} 2022.
\newblock
  \href{http://dx.doi.org/10.17182/hepdata.139717}{\doi{10.17182/hepdata.139717}}.

\end{thebibliography}\endgroup
\cleardoublepage \appendix\section{The CMS Collaboration \label{app:collab}}\begin{sloppypar}\hyphenpenalty=5000\widowpenalty=500\clubpenalty=5000
\cmsinstitute{Yerevan Physics Institute, Yerevan, Armenia}
{\tolerance=6000
A.~Tumasyan\cmsAuthorMark{1}\cmsorcid{0009-0000-0684-6742}
\par}
\cmsinstitute{Institut f\"{u}r Hochenergiephysik, Vienna, Austria}
{\tolerance=6000
W.~Adam\cmsorcid{0000-0001-9099-4341}, J.W.~Andrejkovic, T.~Bergauer\cmsorcid{0000-0002-5786-0293}, S.~Chatterjee\cmsorcid{0000-0003-2660-0349}, K.~Damanakis\cmsorcid{0000-0001-5389-2872}, M.~Dragicevic\cmsorcid{0000-0003-1967-6783}, A.~Escalante~Del~Valle\cmsorcid{0000-0002-9702-6359}, P.S.~Hussain\cmsorcid{0000-0002-4825-5278}, M.~Jeitler\cmsAuthorMark{2}\cmsorcid{0000-0002-5141-9560}, N.~Krammer\cmsorcid{0000-0002-0548-0985}, L.~Lechner\cmsorcid{0000-0002-3065-1141}, D.~Liko\cmsorcid{0000-0002-3380-473X}, I.~Mikulec\cmsorcid{0000-0003-0385-2746}, P.~Paulitsch, J.~Schieck\cmsAuthorMark{2}\cmsorcid{0000-0002-1058-8093}, R.~Sch\"{o}fbeck\cmsorcid{0000-0002-2332-8784}, D.~Schwarz\cmsorcid{0000-0002-3821-7331}, M.~Sonawane\cmsorcid{0000-0003-0510-7010}, S.~Templ\cmsorcid{0000-0003-3137-5692}, W.~Waltenberger\cmsorcid{0000-0002-6215-7228}, C.-E.~Wulz\cmsAuthorMark{2}\cmsorcid{0000-0001-9226-5812}
% ...
\par}
\vskip\cmsinstskip
\dag:~Deceased\\
$^{1}$Also at Yerevan State University, Yerevan, Armenia\\
$^{2}$Also at TU Wien, Vienna, Austria\\
% ...
\end{sloppypar}
\end{document}